\documentclass[12pt,titlepage]{article}

\usepackage{amsfonts}
\usepackage{amsmath}
\usepackage{amssymb}

\begin{document}

\title{Absence of Klein's paradox for massive bosons coupled by nonminimal
vector interactions}
\author{T.R. Cardoso\thanks{%
cardoso@feg.unesp.br }, L.B. Castro\thanks{%
benito@feg.unesp.br }, A.S. de Castro\thanks{%
castro@pq.cnpq.br.} \\
\\
UNESP - Campus de Guaratinguet\'{a}\\
Departamento de F\'{\i}sica e Qu\'{\i}mica\\
12516-410 Guaratinguet\'{a} SP - Brazil}
\date{}
\maketitle

\begin{abstract}
A few properties of the nonminimal vector interactions in the
Duffin-Kemmer-Petiau theory are revised. In particular, it is shown that the
space component of the nonminimal vector interaction plays a peremptory role
for confining bosons whereas its time component contributes to the leakage.
Scattering in a square step potential with proper boundary conditions is
used to show that Klein's paradox does not manifest in the case of a
nonminimal vector coupling.
\end{abstract}

Recently, Boumali \cite{bou} investigated the Duffin-Kemmer-Petiau (DKP)
equation with a sort of nonminimal vector coupling. The very interesting
result of his researching for a step potential is that Klein's paradox can
manifest only for spin-0 particles but not for spin-1 particles. In Ref.
\cite{ccc}, though, it was shown that due to some misconceptions this
conclusion is not reliable. In the present paper we investigate the
properties of the DKP theory with the nonminimal vector coupling
interaction. We use one-dimensional potentials for the sake of simplicity.
It is shown that nonminimal vector potentials have some special features not
displayed by minimal vector potentials. Scattering in a square step
potential is used to show that Klein's paradox never appears in the case of
a nonminimal potential, contrary to what occurs for a minimally coupled
potential \cite{tati}. First observed as a solution of the Dirac equation
\cite{kle} and after that as a solution of the Klein-Gordon equation \cite%
{win}, the famous Klein's paradox consists in a reflection coefficient
exceeding unity under a strong potential. The accepted interpretation for
this phenomenon is that pair production occurs at the potential interface
and that the created particles makes the reflection coefficient greater than
$1$. Of course, Klein's paradox exposes the inadequacy of relativistic
theories as single-particle ones. Furthermore, the present paper shows that
if the space component of the nonminimal potential exceeds its time
component there will be a critical value for the potential strength which
leads to two different possibilities for the waves beyond the potential
interface, either a progressive wave or an evanescent wave, a circumstance
that resembles the nonrelativistic result. Nevertheless, if the space
component of the nonminimal potential does not exceed its time component the
transmission coefficient will never vanish, in sharp contrast to the
nonrelativistic quantum mechanics.

The first-order DKP formalism \cite{pet}-\cite{kem} describes spin-0 and
spin-1 particles and has been used to analyze relativistic interactions of
spin-0 and spin-1 hadrons with nuclei as an alternative to their
conventional second-order Klein-Gordon and Proca counterparts. The onus of
equivalence between the formalisms represented an objection to the DKP
theory for a long time and only recently it was shown that they yield the
same results in the case of minimally coupled vector interactions, on the
condition that one correctly interprets the components of the DKP spinor
\cite{now}-\cite{lun}. The DKP equation for a free boson is given by \cite%
{kem} (with units in which $\hbar =c=1$)%
\begin{equation}
\left( i\beta ^{\mu }\partial _{\mu }-m\right) \psi =0  \label{dkp}
\end{equation}%
\noindent where the matrices $\beta ^{\mu }$\ satisfy the algebra%
\begin{equation}
\beta ^{\mu }\beta ^{\nu }\beta ^{\lambda }+\beta ^{\lambda }\beta ^{\nu
}\beta ^{\mu }=g^{\mu \nu }\beta ^{\lambda }+g^{\lambda \nu }\beta ^{\mu }
\label{beta}
\end{equation}%
\noindent and the metric tensor is $g^{\mu \nu }=\,$diag$\,(1,-1,-1,-1)$.
The algebra expressed by (\ref{beta}) generates a set of 126 independent
matrices whose irreducible representations are a trivial representation, a
five-dimensional representation describing the spin-0 particles and a
ten-dimensional representation associated to spin-1 particles. Indeed, the
Dirac and the DKP equations are special cases of relativistic first-order
equations for arbitrary spin known as Bhabha equation \cite{bha}. The four
square matrices defining the spin, like $\beta ^{\mu }$ in (\ref{dkp}), are
given by representations of the $so(5)$ algebra and furnishes one wave
equation for each spin. The four-dimensional representation $R_{5}\left(
1/2,1/2\right) $, for instance, furnishes the Dirac equation, whilst the
five-dimensional $R_{5}\left( 1,0\right) $ and the ten-dimensional $%
R_{5}\left( 1,1\right) $ representations furnish the DKP equation for spin-0
and spin-1, respectively \cite{nie}. The DKP spinor has an excess of
components and the theory has to be supplemented by an equation which allows
to eliminate the redundant components. That constraint equation is obtained
by multiplying the DKP equation by $1-\beta ^{0}\beta ^{0}$, namely%
\begin{equation}
i\beta ^{j}\beta ^{0}\beta ^{0}\partial _{j}\psi =m\left( 1-\beta ^{0}\beta
^{0}\right) \psi ,\quad j\text{ \ runs from 1 to 3}  \label{vin1}
\end{equation}%
This constraint equation expresses three (four) components of the spinor by
the other two (six) components and their space derivatives in the scalar
(vector) sector so that the superfluous components disappear and there only
remain the physical components of the DKP theory. The second-order
Klein-Gordon and Proca equations are obtained when one selects the spin-0
and spin-1 sectors of the DKP theory. A well-known conserved four-current is
given by
\begin{equation}
J^{\mu }=\bar{\psi}\beta ^{\mu }\psi   \label{corrente}
\end{equation}%
\noindent where the adjoint spinor $\bar{\psi}=\psi ^{\dagger }\eta ^{0}$,
with $\eta ^{0}=2\beta ^{0}\beta ^{0}-1$ in such a way that $\left( \eta
^{0}\beta ^{\mu }\right) ^{\dagger }=\eta ^{0}\beta ^{\mu }$ (the matrices $%
\beta ^{\mu }$ are Hermitian with respect to $\eta ^{0}$). The time
component of this current is not positive definite but it may be interpreted
as a charge density.

With the introduction of interactions, the DKP equation can be written as%
\begin{equation}
\left( i\beta ^{\mu }\partial _{\mu }-m-\mathbb{V}\right) \psi =0
\label{dkp2}
\end{equation}%
where the more general potential matrix $\mathbb{V}$ is written in terms of
25 (100) linearly independent matrices pertinent to the
five(ten)-dimensional irreducible representation associated to the scalar
(vector) sector. In the presence of interactions $J^{\mu }$ satisfies the
equation%
\begin{equation}
\partial _{\mu }J^{\mu }+i\bar{\psi}\left( \mathbb{V}-\eta ^{0}\mathbb{V}%
^{\dagger }\eta ^{0}\right) \psi =0  \label{corrent2}
\end{equation}%
Thus, if $\mathbb{V}$ is Hermitian with respect to $\eta ^{0}$ then the
four-current will be conserved. The potential matrix $\mathbb{V}$ can be
written in terms of well-defined Lorentz structures. For the spin-0 sector
there are two scalar, two vector and two tensor terms \cite{gue}, whereas
for the spin-1 sector there are two scalar, two vector, a pseudoscalar, two
pseudovector and eight tensor terms \cite{vij}. The tensor terms have been
avoided in applications because they furnish noncausal effects \cite{gue}-%
\cite{vij}. By considering only the nonminimal vector terms, $\mathbb{V}$ is
in the form%
\begin{equation}
\mathbb{V}=i[P,\beta ^{\mu }]A_{\mu }  \label{pot}
\end{equation}%
where $P$ is a projection operator ($P^{2}=P$ and $P^{\dagger }=P$) in such
a way that $\bar{\psi}P\psi $ behaves as a scalar and $\bar{\psi}[P,\beta
^{\mu }]\psi $ behaves like a vector. $A_{\mu }$ is the four-vector
potential function. At this point it is also worthwhile to note that this
matrix potential leads to a conserved four-current but the same does not
happen if instead of $i[P,\beta ^{\mu }]$ one uses either $P\beta ^{\mu }$
or $\beta ^{\mu }P$.

For the case of spin 0, we use the representation for the $\beta ^{\mu }$\
matrices given by \cite{ned1}-\cite{ned3}%
\begin{equation}
\beta ^{0}=%
\begin{pmatrix}
\theta & \overline{0} \\
\overline{0}^{T} & \mathbf{0}%
\end{pmatrix}%
,\quad \beta ^{i}=%
\begin{pmatrix}
\widetilde{0} & \rho _{i} \\
-\rho _{i}^{T} & \mathbf{0}%
\end{pmatrix}%
,\quad i=1,2,3  \label{rep}
\end{equation}%
\noindent where%
\begin{eqnarray}
\ \theta &=&%
\begin{pmatrix}
0 & 1 \\
1 & 0%
\end{pmatrix}%
,\quad \rho _{1}=%
\begin{pmatrix}
-1 & 0 & 0 \\
0 & 0 & 0%
\end{pmatrix}
\notag \\
&&  \label{rep2} \\
\rho _{2} &=&%
\begin{pmatrix}
0 & -1 & 0 \\
0 & 0 & 0%
\end{pmatrix}%
,\quad \rho _{3}=%
\begin{pmatrix}
0 & 0 & -1 \\
0 & 0 & 0%
\end{pmatrix}
\notag
\end{eqnarray}%
\noindent $\overline{0}$, $\widetilde{0}$ and $\mathbf{0}$ are 2$\times $3, 2%
$\times $2 and 3$\times $3 zero matrices, respectively, while the
superscript T designates matrix transposition. Here the projection operator
can be written as \cite{gue}
\begin{equation}
P=\,\frac{1}{3}\left( \beta ^{\mu }\beta _{\mu }-1\right) =\text{diag}%
\,(1,0,0,0,0)  \label{proj}
\end{equation}%
In this case $P$ picks out the first component of the DKP spinor. The
five-component spinor can be written as $\psi ^{T}=\left( \psi _{1},...,\psi
_{5}\right) $ in such a way that the DKP equation for a boson constrained to
move along the $x$-axis decomposes into
\begin{equation*}
\left( D_{0}^{\left( -\right) }D_{0}^{\left( +\right) }-D_{1}^{\left(
-\right) }D_{1}^{\left( +\right) }+m^{2}\right) \psi _{1}=0
\end{equation*}%
\begin{equation}
D_{0}^{\left( +\right) }\psi _{1}=-im\psi _{2},\quad D_{1}^{\left( +\right)
}\psi _{1}=-im\psi _{3}  \label{DKP3}
\end{equation}%
\begin{equation*}
\psi _{4}=\psi _{5}=0
\end{equation*}%
where%
\begin{equation}
D_{\mu }^{\left( \pm \right) }=\partial _{\mu }\pm A_{\mu }  \label{dzao}
\end{equation}%
Here we return to (\ref{corrente}), which we rewrite as%
\begin{equation}
J^{0}=2\,\text{Re}\left( \psi _{2}^{\ast }\psi _{1}\right) ,\quad J^{1}=-2\,%
\text{Re}\left( \psi _{3}^{\ast }\psi _{1}\right) ,\quad J^{2}=J^{3}=0
\label{corrente3}
\end{equation}%
If the terms in the potential $\mathbb{V}$ are time-independent, one can
write $\psi (x,t)=\varphi (x)\exp (-iEt)$ in such a way that the
time-independent DKP equation splits into%
\begin{equation*}
\left( \frac{d^{2}}{dx^{2}}+\kappa ^{2}\right) \varphi _{1}=0
\end{equation*}%
\begin{equation*}
\varphi _{2}=\frac{E+iA_{0}}{m}\,\varphi _{1}
\end{equation*}%
\begin{equation}
\varphi _{3}=\frac{i}{m}\left( \frac{d}{dx}+A_{1}\right) \varphi _{1}
\label{dkp4}
\end{equation}%
\begin{equation*}
\varphi _{4}=\varphi _{5}=0
\end{equation*}%
where%
\begin{equation}
\kappa ^{2}=E^{2}-m^{2}+A_{0}^{2}-A_{1}^{2}+\frac{dA_{1}}{dx}  \label{k}
\end{equation}%
For this time-independent problem, $J^{\mu }$ has the components
\begin{equation}
J^{0}=\frac{2E}{m}\,|\varphi _{1}|^{2},\;J^{1}=\frac{2}{m}\,\text{Im}\left(
\varphi _{1}^{\ast }\frac{d\varphi _{1}}{dx}\right)  \label{corrente4}
\end{equation}%
Since $J^{\mu }$ is not time dependent, $\varphi $ describes a stationary
state.

The one-dimensional square step potential is expressed as
\begin{equation}
A_{\mu }=\theta \left( x\right) c_{\mu }V  \label{pot1}
\end{equation}%
\noindent where $c_{\mu }$ are dimensionless and positive coupling constants
constrained by $c_{0}+c_{1}=1$, $\theta \left( x\right) $ denotes the
Heaviside step function and $V>0$ is the height of the step. For $x<0$ the
DKP equation has the solution
\begin{equation}
\varphi \left( x\right) =\varphi _{+}e^{+ikx}+\varphi _{-}e^{-ikx}
\label{sol z<0}
\end{equation}%
\noindent where%
\begin{equation}
\varphi _{\pm }^{T}=\frac{a_{\pm }}{\sqrt{2}}\left( 1,\frac{E}{m},\mp \frac{k%
}{m},0,0\right)  \label{sol2}
\end{equation}%
\noindent and%
\begin{equation}
k=\sqrt{E^{2}-m^{2}}  \label{kkk}
\end{equation}%
For $\left\vert E\right\vert >m$, the solution expressed by (\ref{sol z<0})
and (\ref{sol2}) describes plane waves propagating on both directions of the
$x$-axis with the group velocity $v_{g}=dE/dk$ equal to the classical
velocity. If we choose incident particles on the potential barrier $\left(
E>m\right) $, $\varphi _{+}\exp (+ikx)$ will describe incident particles ($%
v_{g}=+k/E>0$), whereas $\varphi _{-}\exp (-ikx)$ will describe reflected
particles ($v_{g}=-k/E<0$). The flux related to the current $J^{\mu }$,
corresponding to $\varphi $ given by (\ref{sol z<0}), is expressed as
\begin{equation}
J^{1}=\frac{k}{m}\left( \left\vert a_{+}\right\vert ^{2}-\left\vert
a_{-}\right\vert ^{2}\right)  \label{j1}
\end{equation}%
\noindent Note that the relation $J^{1}=J^{0}\,v_{g}$ maintains for the
incident and reflected waves, since%
\begin{equation}
J_{\pm }^{0}=\frac{E}{m}\left\vert a_{\pm }\right\vert ^{2}  \label{j2}
\end{equation}%
\noindent On the other hand, for $x>0$ one should have $v_{g}\geq 0$ in such
a way that the solution in this region of space describes an evanescent wave
or a progressive wave running away from the potential interface. The general
solution has the form
\begin{equation}
\varphi _{\text{t}}\left( x\right) =\left( \varphi _{\text{t}}\right)
_{+}e^{+iqx}+\left( \varphi _{\text{t}}\right) _{-}e^{-iqx}  \label{sol z0}
\end{equation}%
\noindent where%
\begin{equation}
\left( \varphi _{\text{t}}\right) _{\pm }^{T}=\frac{b_{\pm }}{\sqrt{2}}%
\left( 1,\frac{E+ic_{0}V}{m},\frac{\mp q+ic_{1}V}{m},0,0\right)  \label{sol3}
\end{equation}%
\noindent and%
\begin{equation}
q=\sqrt{k^{2}+\left( c_{0}-c_{1}\right) V^{2}}  \label{q}
\end{equation}%
Due to the twofold possibility of signs for the energy of a stationary
state, the solution involving $b_{-}$ can not be ruled out a priori. As a
matter of fact, this term may describe a progressive wave with negative
energy and phase velocity $v_{ph}=|E|/q>0$ (see, e.g. \cite{tati}). In other
words, the solution $\left( \varphi _{\text{t}}\right) _{-}\exp \left(
-iqx\right) $ with $q\in \mathbb{R}$ reveals a signature of Klein's paradox.
One can readily envisage that two different classes of solutions can be
distinguished:

\begin{itemize}
\item Class A. With $c_{1}>c_{0}$ for $V<V^{c}$, where
\begin{equation}
V^{c}=\sqrt{\frac{E^{2}-m^{2}}{c_{1}-c_{0}}}  \label{vc}
\end{equation}%
or with $c_{1}\leq c_{0}$ for all $V$, one has $q\in\mathbb{R}$, and the
solution describing a plane wave propagating in the positive direction of
the $x$-axis with the group velocity $v_{g}=q/E$ is possible only if $%
b_{-}=0 $. In this case the components of the current are given by

\begin{equation}
J^{0}=\frac{E}{m}\left\vert b_{+}\right\vert ^{2},\;J^{1}=\frac{q}{m}%
\left\vert b_{+}\right\vert ^{2}  \label{c11}
\end{equation}

\item Class B. With $c_{1}>c_{0}$ for $V>V^{c}$ one has that $q=\pm
i\left\vert q\right\vert $, and (\ref{sol z0}) with $b_{\mp }=0$ describes
an evanescent wave. The solution satisfying the requirement of finiteness at
infinity requires $b_{\mp }=0$. In this case\bigskip
\begin{equation}
J^{0}=\frac{E}{m}e^{-2\left\vert q\right\vert x}\left\vert b_{\pm
}\right\vert ^{2},\;J^{1}=0  \label{c2}
\end{equation}

Incidentally, the solution involving $b_{-}$ is identical to the solution
involving $b_{+}$, so we consider $b_{-}=0$.
\end{itemize}

Note that there is no reason to require that the either the spinor and or
its derivative are continuous across finite discontinuities of the
potential. A careful analysis reveals, though, that proper matching
conditions follow from the differential equations obeyed by the spinor
components. Only the first component of the spinor satisfies the
second-order Klein-Gordon-like equation, so that $\varphi _{1}$ and its
first derivative are continuous even the potential suffers finite
discontinuities. In this case of a discontinuous potential, $\varphi _{2}$
(if $A_{0}\neq 0$) and $\varphi _{3}$ (if $A_{1}\neq 0$) are discontinuous
but $J^{0}$ and $J^{1}$ are not. A possible discontinuity of $J^{0}$ would
not matter if it is to be interpreted as a charge density but $J^{1}$
(involving $\varphi _{1}^{\ast }\,d\varphi _{1}/dx$) should be continuous in
a stationary regime. The demand for continuity of $\varphi _{1}$ and $%
d\varphi _{1}/dx$ at $x=0$ fixes the wave amplitudes in terms of the
amplitude of the incident wave, viz.
\begin{equation}
\frac{a_{-}}{a_{+}}=\left\{
\begin{array}{c}
\frac{k-q}{k+q} \\
\\
\frac{\left( k-i|q|\right) ^{2}}{k^{2}+|q|^{2}}%
\end{array}%
\begin{array}{c}
\text{\textrm{for} \textrm{the class A}} \\
\\
\text{\textrm{for} \textrm{the class B}}%
\end{array}%
\right.  \label{12}
\end{equation}%
\begin{equation}
\frac{b_{+}}{a_{+}}=\left\{
\begin{array}{c}
\frac{2k}{k+q} \\
\\
\frac{2k\left( k-i|q|\right) }{k^{2}+|q|^{2}}%
\end{array}%
\begin{array}{c}
\text{\textrm{for} \textrm{the class A}} \\
\\
\text{\textrm{for} \textrm{the class B}}%
\end{array}%
\right.  \label{13}
\end{equation}

Now we focus attention on the calculation of the reflection ($R$) and
transmission ($T$) coefficients. The reflection (transmission) coefficient
is defined as the ratio of the reflected (transmitted) flux to the incident
flux. Since $\partial J^{0}/\partial t=0$ for stationary states, one has
that $J^{1}$ is independent of $x$. This fact implies that
\begin{equation}
R=\left\{
\begin{array}{c}
\left( \frac{k-q}{k+q}\right) ^{2} \\
\\
1%
\end{array}%
\begin{array}{c}
\text{\textrm{for} \textrm{the class A}} \\
\\
\text{\textrm{for} \textrm{the class B}}%
\end{array}%
\right.  \label{15}
\end{equation}%
\begin{equation}
T=\left\{
\begin{array}{c}
\frac{4kq}{(k+q)^{2}} \\
\\
0%
\end{array}%
\begin{array}{c}
\text{\textrm{for} \textrm{the class A}} \\
\\
\text{\textrm{for} \textrm{the class B}}%
\end{array}%
\right.  \label{16}
\end{equation}

\noindent For all the classes one has $R+T=1$ as should be expected for a
conserved quantity. Note that the charge density in (\ref{c11}) and (\ref{c2}%
) is always a positive quantity and so is $J^{1}$ in (\ref{c11}). This means
that the scattered waves describe particles and not antiparticles, then
Klein's paradox never manifests.

For $c_{1}>c_{0}$ the transmission coefficient vanishes for a potential
strength $V$ greater than the cutoff potential $V^{c}$. In fact, the mixed
step potential behaves effectively as a ascending step and a similar
situation occurs in nonrelativistic quantum mechanics. The uncertainty in
the position beyond the potential boundary for $V>V^{c}$ can be obtained
from (\ref{c2}), namely
\begin{equation}
\Delta x=\frac{1}{2|q|}=\frac{1}{2\sqrt{m^{2}+\left( c_{1}-c_{0}\right)
V^{2}-E^{2}}}  \label{Dx}
\end{equation}%
From this last result one can see that the penetration of the boson into the
region $x>0$ will shrink without limit as $V$ increases. At first glance it
seems that the uncertainty principle dies away provided such a principle
implies that it is impossible to localize a particle into a region of space
less than half of its Compton wavelength (see, e.g., \cite{gre} and \cite%
{str}). This apparent contradiction can be remedied by recurring to the
concepts of effective mass and effective Compton wavelength. Indeed, Eq. (%
\ref{Dx}) suggests that we can define the effective mass as%
\begin{equation}
m_{\mathtt{eff}}=\sqrt{m^{2}+\left( c_{1}-c_{0}\right) V^{2}}  \label{meff}
\end{equation}%
in such a way that%
\begin{equation}
\Delta x=\frac{1}{2\sqrt{m_{\mathtt{eff}}^{2}-E^{2}}}  \label{Dxef}
\end{equation}%
The effective mass clearly indicates that this kind of potential couples to
the mass of the boson and consequently it couples to the positive-energy
component of the spinor in the same way it couples to the negative-energy
component. This amounts to say that Klein's paradox does not appear. It is
seen that the minimum uncertainty is $\left( \Delta x\right) _{\min
}=1/\left( 2m_{\mathtt{eff}}\right) $ in the limit as $V\rightarrow \infty $%
. Therefore, for obtaining a result consistent with the uncertainty
principle it is necessary to use the effective Compton wavelength defined as
$\lambda _{\mathtt{eff}}=1/m_{\mathtt{eff}}$ so that the minimum uncertainty
consonant with the uncertainty principle is given by $\lambda _{\mathtt{eff}%
}/2$. It means that the localization of the boson does not require any
minimum value in order to ensure the single-particle interpretation of the
DKP equation.

As for $c_{1}\leq c_{0}$, however, there is no cutoff potential. This is a
result that runs counter our conceptions drawn from the nonrelativistic
quantum mechanics. For $c_{1}=c_{0}$ the half-and-half mixed step potential
is transparent ($T=1$ for all $V$), and for $c_{1}<c_{0}$ the mixed step
presents a transmission coefficient that goes as%
\begin{equation}
T\rightarrow \frac{4}{V}\,\sqrt{\frac{E^{2}-m^{2}}{c_{0}-c_{1}}}
\label{Tgoes}
\end{equation}%
as $V\rightarrow \infty $. Those strange facts occur because the space
component of the step potential behaves as an ascending step whereas its
time component behaves as a descending step. For $c_{1}=c_{0}$, although $%
\varphi _{+}\neq \left( \varphi _{\text{t}}\right) _{+}$, effects due to the
time and the space components cancel each other as far as the transmission
coefficient is regarded. That is to say, the effective potential behaves as
a transparent potential. For $c_{1}<c_{0}$ the tendency to a descending step
dominates so that the mixed step potential effectively behaves as a
descending step. Note that the reflection and transmission coefficients are
the same for a wave incident from the right as for a wave incident from the
left.

To conclude, we have shown minimal and nonminimal vector interactions in the
DKP theory behave quite diversely. In particular, nonminimal vector
interactions have no counterparts in the Klein-Gordon theory. Nonminimal
vector interactions have the very same effects on both particles and
antiparticles and so they might be useful for boson-confining models.
Scattering in a square step potential clearly shows that Klein's paradox,
present in the case of a minimal coupling \cite{tati}, is absent in the case
of a nonminimal coupling. An apparent contradiction with the uncertainty
principle has been cured by introducing the concepts of effective mass and
effective Compton wavelength. When the space component of the nonminimal
potential does not exceed its time component, the transmission coefficient
is different from zero even if the height of the step potential is extremely
high. That odd result has been endorsed by observing the behaviour of the
effective potential.

We have talking about the spin-0 sector of the DKP theory but the state of
affairs for the step potential is not different for the spin-1 sector as one
can see in Appendix A.

\bigskip

\bigskip \bigskip

\bigskip

\noindent \textbf{Appendix A}

For the case of spin 1, the $\beta ^{\mu }$\ matrices are \cite{ned3}, \cite%
{ned2}%
\begin{equation}
\beta ^{0}=%
\begin{pmatrix}
0 & \overline{0} & \overline{0} & \overline{0} \\
\overline{0}^{T} & \mathbf{0} & \mathbf{I} & \mathbf{0} \\
\overline{0}^{T} & \mathbf{I} & \mathbf{0} & \mathbf{0} \\
\overline{0}^{T} & \mathbf{0} & \mathbf{0} & \mathbf{0}%
\end{pmatrix}%
,\;\beta ^{i}=%
\begin{pmatrix}
0 & \overline{0} & e_{i} & \overline{0} \\
\overline{0}^{T} & \mathbf{0} & \mathbf{0} & -is_{i} \\
-e_{i}^{T} & \mathbf{0} & \mathbf{0} & \mathbf{0} \\
\overline{0}^{T} & -is_{i} & \mathbf{0} & \mathbf{0}%
\end{pmatrix}
\label{betaspin1}
\end{equation}%
\noindent where $s_{i}$ are the 3$\times $3 spin-1 matrices $\left(
s_{i}\right) _{jk}=-i\varepsilon _{ijk}$, $e_{i}$ are the 1$\times $3
matrices $\left( e_{i}\right) _{1j}=\delta _{ij}$ and $\overline{0}=%
\begin{pmatrix}
0 & 0 & 0%
\end{pmatrix}%
$, while $\mathbf{I}$ and $\mathbf{0}$ designate the 3$\times $3 unit and
zero matrices, respectively. In this representation
\begin{equation}
P=\,\beta ^{\mu }\beta _{\mu }-2=\text{diag}\,(1,1,1,1,0,0,0,0,0,0)
\label{p1}
\end{equation}%
\textit{i.e.}, $P$ projects out the four upper components of the DKP spinor.
\noindent With the spinor written as $\psi ^{T}=\left( \psi _{1},...,\psi
_{10}\right) $, and partitioned as%
\begin{equation*}
\psi _{I}^{\left( +\right) }=\left(
\begin{array}{c}
\psi _{3} \\
\psi _{4}%
\end{array}%
\right) ,\quad \psi _{I}^{\left( -\right) }=\psi _{5}
\end{equation*}%
\begin{equation}
\psi _{II}^{\left( +\right) }=\left(
\begin{array}{c}
\psi _{6} \\
\psi _{7}%
\end{array}%
\right) ,\quad \psi _{II}^{\left( -\right) }=\psi _{2}  \label{part}
\end{equation}%
\begin{equation*}
\psi _{III}^{\left( +\right) }=\left(
\begin{array}{c}
\psi _{10} \\
-\psi _{9}%
\end{array}%
\right) ,\quad \psi _{III}^{\left( -\right) }=\psi _{1}
\end{equation*}%
the one-dimensional DKP equation can be expressed in the form
\begin{equation*}
\left( D_{0}^{\left( \mp \right) }D_{0}^{\left( \pm \right) }-D_{1}^{\left(
\mp \right) }D_{1}^{\left( \pm \right) }+m^{2}\right) \psi _{I}^{\left( \pm
\right) }=0
\end{equation*}%
\begin{equation}
D_{0}^{\left( \pm \right) }\psi _{I}^{\left( \pm \right) }=-im\psi
_{II}^{\left( \pm \right) },\quad D_{1}^{\left( \pm \right) }\psi
_{I}^{\left( \pm \right) }=-im\psi _{III}^{\left( \pm \right) }  \label{DKp3}
\end{equation}%
\begin{equation*}
\psi _{8}=0
\end{equation*}%
where $D_{\mu }^{\left( \pm \right) }$ is again given by (\ref{dzao}). In
addition, expressed in terms of (\ref{part}) the current can be written as
\begin{equation*}
J^{0}=2\,\text{Re}\left( \psi _{II}^{\left( +\right) \dagger }\psi
_{I}^{\left( +\right) }+\psi _{II}^{\left( -\right) \dagger }\psi
_{I}^{\left( -\right) }\right)
\end{equation*}%
\begin{equation}
J^{1}=-2\,\text{Re}\left( \psi _{III}^{\left( +\right) \dagger }\psi
_{I}^{\left( +\right) }+\psi _{III}^{\left( -\right) \dagger }\psi
_{I}^{\left( -\right) }\right)  \label{CUR}
\end{equation}%
\begin{equation*}
J^{2}=J^{3}=0
\end{equation*}%
Meanwhile the time-independent DKP equation decomposes into%
\begin{equation*}
\left( \frac{d^{2}}{dx^{2}}+k_{\pm }^{2}\right) \phi _{I}^{\left( \pm
\right) }=0
\end{equation*}%
\begin{equation}
\phi _{II}^{\left( \pm \right) }=\frac{1}{m}\left( E\pm iA_{0}\right) \,\phi
_{I}^{\left( \pm \right) }  \label{spin1-ti}
\end{equation}%
\begin{equation*}
\phi _{III}^{\left( \pm \right) }=\frac{i}{m}\left( \frac{d}{dx}\pm
A_{1}\right) \phi _{I}^{\left( \pm \right) }
\end{equation*}%
where%
\begin{equation}
k_{\pm }^{2}=E^{2}-m^{2}+A_{0}^{2}-A_{1}^{2}\pm \frac{dA_{1}}{dx}  \label{K}
\end{equation}%
Now the components of the four-current are
\begin{equation*}
J^{0}=\frac{2}{m}E\left( |\phi _{I}^{\left( +\right) }|^{2}+|\phi
_{I}^{\left( -\right) }|^{2}\right)
\end{equation*}%
\begin{equation}
J^{1}=\frac{2}{m}\text{Im}\left( \phi _{I}^{\left( +\right) \dagger }\,\frac{%
d\phi _{I}^{\left( +\right) }}{dx}+\phi _{I}^{\left( -\right) \dagger }\,%
\frac{d\phi _{I}^{\left( -\right) }}{dx}\right)  \label{CUR2}
\end{equation}%
From (\ref{spin1-ti})-(\ref{K}) one sees that the solution for the spin-1
sector consists in searching solutions for two Klein-Gordon-like equations,
owing to the term $dA_{1}/dx$ in (\ref{K}). For the square step potential
given by (\ref{pot1}), because $dA_{1}/dx=0$ for $x\neq 0$ one has that $%
\phi _{I}^{\left( +\right) }$ and $\phi _{I}^{\left( -\right) }$ obey the
same equation so that the solution in the region $x<0$ can be written as%
\begin{equation}
\phi _{I}^{\left( \pm \right) }=A_{+}^{\left( \pm \right)
}e^{+ikx}+A_{-}^{\left( \pm \right) }e^{-ikx}
\end{equation}%
where $k$ is once again given by (\ref{kkk}), and $A_{\pm }^{\left( \pm
\right) }$ is defined in terms of the arbitrary amplitudes $\alpha _{\pm }$,
$\beta _{\pm }$ and $\gamma _{\pm }$ as
\begin{equation}
A_{\pm }^{\left( +\right) }=\left(
\begin{array}{c}
\alpha _{\pm } \\
\beta _{\pm }%
\end{array}%
\right) ,\quad A_{\pm }^{\left( -\right) }=\gamma _{\pm }
\end{equation}%
Defining%
\begin{equation}
a_{\pm }=\sqrt{2\left( |\alpha _{\pm }|^{2}+|\beta _{\pm }|^{2}+|\gamma
_{\pm }|^{2}\right) }
\end{equation}%
it follows that the components of the current can be written in the same
form as (\ref{j1}) and (\ref{j2}). A similar procedure for the region $x>0$
allows one to obtain the results (\ref{12})-(\ref{Tgoes}).

\bigskip

\noindent \textbf{Acknowledgments}

This work was supported in part by means of funds provided by CAPES and CNPq.

\end{document}